\documentstyle[12pt]{article}

\newcommand{\bra}[1]{\left\langle#1\right|}
\newcommand{\ket}[1]{\left|#1\right\rangle}
\newcommand{\mtx}[2]{\left(\begin{array}{#1}#2\end{array}\right)}
\newcommand{\tr}{\mbox{Tr}\,} 

\begin{document}

\begin{center}

\bigskip {\Large Entanglement of a Pair of Quantum Bits}\\

\bigskip

Scott Hill and William K.~Wootters\\

\bigskip

{\small{\sl

Department of Physics, Williams College, Williamstown MA 01267}}
\vspace{2mm}

\end{center} \subsection*{\centering Abstract} {The ``entanglement of
formation'' of a mixed state of a bipartite quantum system can be defined
in terms of the number of pure singlets needed to create the state with
no further transfer of quantum information.  We find an exact formula
for the entanglement of formation for all mixed  states of two qubits
having no more than two non-zero eigenvalues,  and we report evidence
suggesting that the formula is valid for all states of this system.}

\vfill

PACS numbers: 03.65.Bz, 89.70.+c\vfill

\newpage

Entanglement is the potential of quantum states to exhibit correlations
that cannot be accounted for classically.  For decades, entanglement has
been the focus of much work in the foundations of quantum mechanics,
being associated particularly with quantum nonseparability and the
violation of Bell's inequalities \cite{Bell}.  In recent years, however,
it has begun to be viewed also as a potentially useful resource.  The
predicted capabilities of a quantum computer, for example, rely crucially
on entanglement \cite{computer}, and a proposed quantum cryptographic
scheme achieves security by converting shared entanglement into a shared
secret key \cite{crypto}.  For both theoretical and potentially practical
reasons, it has become interesting to quantify entanglement, just as we
quantify other resources such as energy and information.  In this letter
we adopt a recently proposed quantitative definition of entanglement and
derive an explicit formula for the entanglement of a large class of
states of a pair of binary quantum systems (qubits).

The simplest kind of entangled system is a pair of qubits in a {\em pure}
but non-factorizable state.  A pair of spin-${1\over 2}$ particles in the
singlet state  \mbox{${1\over
\sqrt{2}}(|\uparrow\downarrow\rangle~-~|\downarrow\uparrow\rangle)$} is
perhaps the most familiar example, but one can also consider more general
states such as
$\alpha|\uparrow\downarrow\rangle+\beta|\downarrow\uparrow\rangle$,
which may be less entangled.  For any bipartite system in a pure state,
Bennett {\em et al.} \cite{pure} have shown that it is reasonable to
define the entanglement of the system as the von Neumann entropy of
either of its two parts.  That is, if $|\psi\rangle$ is the state of the
whole system, the entanglement can be defined as $E(\psi) = -\tr \rho
\log_2 \rho$, where $\rho$ is the partial trace of
$|\psi\rangle\langle\psi|$ over either of the two subsystems.  (It does
not matter which subsystem one traces over; the result is the same either
way.)  What Bennett {\em et al.} showed specifically is the following.
Consider $n$ pairs, each in the state $|\psi\rangle$.  Let an observer
Alice hold one member of each pair and let Bob, whom we imagine to be
spatially separated from Alice, hold the other.  Then if $|\psi\rangle$
has $E$ ``ebits'' of entanglement according to the above definition, the
$n$ pairs can be reversibly converted by purely local operations and
classical communication into $m$ pairs of qubits in the singlet state,
where $m/n$ approaches $E$ for large $n$ and the fidelity of the
conversion approaches 100\%.  This interconvertibility is strong
justification for the above definition of $E$ and characterizes it
uniquely.

It is somewhat harder to define the entanglement of mixed states
\cite{vedral}, though  again one can use the singlet as the basic unit of
entanglement and relate the given mixed state to singlets.  The new
feature in the case of mixed states is that the number of singlets
required to {\em create} the state is not necessarily the same as the
number of singlets one can {\em extract} from the state \cite{distill}.
In this paper we focus on the former quantity, which leads to the concept
of ``entanglement of formation.''  In mathematical terms the entanglement
of formation is defined as follows \cite{formation}. Given a mixed state
$\rho$ of two quantum systems $A$ and $B$, consider all possible ways of
expressing $\rho$ as an ensemble of pure states. That is, we consider
states $|\psi_i\rangle$ and associated probabilities $p_i$ such that

\begin{equation} \rho = \sum_i p_i |\psi_i \rangle \langle \psi_i |.
\end{equation} The entanglement of formation of $\rho$, $E(\rho)$, is
defined as the minimum, over all such ensembles, of the average
entanglement of the pure states making up the ensemble:

\begin{equation} E = \hbox{min \,}\sum_i p_i E(\psi_i). \label{Emin}
\end{equation} Entanglement of formation has the satisfying property that
it is zero if and only if the state in question can be expressed as a
mixture of product states. For ease of expression, we will refer to the
entanglement of formation simply as ``entanglement.''

Peres \cite{Peres} and Horodecki {\em et al.} \cite{Horodecki} have found
elegant characterizations of states with zero and non-zero $E$, and
Bennett {\em et al.} \cite{formation} have determined the value of $E$
for mixtures of Bell states. (These are a particular set of orthogonal,
completely entangled states of two qubits; we will refer to other sets of
such  states as ``generalized Bell states.'')   But the value of $E$ for
most states, even of two qubits, is not known, and in fact it has not
been evident that one can even express $E$ in closed form as a function
of the density matrix.  The exact formula we derive in this letter is
proved for all density matrices of two qubits having  only two non-zero
eigenvalues, but it appears likely that it applies to {\em all} states of
this  system.

Our starting point is a curious and useful fact about the pure states of
a pair of qubits.  For such a system, we define a ``magic basis''
consisting of the following four states (they are the Bell states with
particular phases) \cite{formation}: \begin{equation}\begin{array}{l}
\ket{e_1}=\frac
12\left(\ket{\uparrow\uparrow}+\ket{\downarrow\downarrow}\right)\\
\ket{e_2}=\frac 12
i\left(\ket{\uparrow\uparrow}-\ket{\downarrow\downarrow}\right)\\
\ket{e_3}=\frac 12
i\left(\ket{\uparrow\downarrow}+\ket{\downarrow\uparrow}\right)\\
\ket{e_4}=\frac
12\left(\ket{\uparrow\downarrow}-\ket{\downarrow\uparrow}\right)
\end{array}\end{equation} where we have used spin-${1\over 2}$ notation
for definiteness.  When a pure state $\ket{\psi}$ is written in this
particular basis, as $|\psi\rangle = \sum_i \alpha_i |e_i\rangle$, its
entanglement can be expressed in a remarkably simple way \cite{formation}
in terms of the components $\alpha_i$:  Define the  function

\begin{equation} {\cal E}(x) = H\Bigl({1\over 2} + {1\over
2}\sqrt{1-x^2}\Bigr) \hspace{1cm}\mbox{for } 0 \le x \le 1, \label{eps}
\end{equation} where $H$ is the binary entropy function $H(x) = -[x\log_2
x + (1-x)\log_2 (1-x)]$.  Then the entanglement of $\ket{\psi}$ is

\begin{equation} E(\psi) = {\cal E}(C(\psi)),\label{Eeq} \end{equation}
where $C$ is defined by

\begin{equation} C(\psi) = | \sum_i \alpha^2_i |. \label{concurrence}
\end{equation} The quantity $C$, like $E$ for this system, ranges from
zero to one, and it is monotonically related to $E$, so that $C$ is a
kind of measure of entanglement in its own right.  It is sufficiently
useful that we give it its own name: concurrence.  As we look for a
pure-state ensemble with minimum {\em average} entanglement for a given
mixed state, our plan will be to look for a set of states that all have
the {\em same} entanglement, which is to say that they all have the same
concurrence.

Two other facts about the magic basis are worth highlighting.  (i) The
set of states whose density matrices are {\em real} when expressed in the
magic basis  is the same as the set of  mixtures of generalized Bell
states (Horodecki {\em et al.} have called such mixtures ``T~states''
\cite{Horodecki2}). (ii) The set of unitary transformations that are real
when expressed in the magic basis (or real except  for an overall phase
factor) is the same as the set of transformations that act independently
on the two qubits.

It happens that our formula for $E$ is conveniently expressed in terms of
a  matrix $R$, which is a function of $\rho$ defined by the equation

\begin{equation} R(\rho) = \sqrt{\sqrt{\rho}\rho^*\sqrt{\rho}}.\label{R}
\end{equation}  Here $\rho^*$ is the complex conjugate of $\rho$ when it
is expressed in the magic basis; that is,  $\rho^* = \sum_{ij}
|e_i\rangle\langle e_j|\rho|e_i\rangle\langle e_j|$. To get some sense of
the meaning of $R$, note that Tr\,$R$, ranging from 0 to 1,  is a measure
of the ``degree of equality'' \cite{Bures} between $\rho$ and $\rho^*$,
which in turn measures how nearly $\rho$ approximates a mixture of
generalized Bell states.  Note also that the eigenvalues of $R$ are
invariant under local unitary transformations of the separate qubits, a
fact that makes these  eigenvalues particularly eligible to be part of a
formula for entanglement,  since entanglement must also be invariant
under such transformations.  We now state our main result.

{\em Theorem}.  Let $\rho$ be any density matrix of two qubits having no
more than two non-zero eigenvalues.  Let $\lambda_{max}$ be the largest
eigenvalue of $R(\rho)$.   Then the entanglement of formation of $\rho$
is given by  \begin{equation} E(\rho) = {\cal E}(c); \hskip 2pc c =
{\hbox{max}\,}(0, 2\lambda_{max} - {\mbox{Tr}\,}R).\label{E4}
\end{equation} (The quantity $c$ can thus be called the concurrence of
the density matrix $\rho$. If $\rho$ is pure, then $c$ reduces to the
concurrence defined in Eq.~(\ref{concurrence}).)

{\em Proof}. Let $|v_1\rangle$ and $|v_2\rangle$ be the two eigenvectors
of $\rho$ corresponding to its two non-zero eigenvalues. Define the $2
\times 2$ matrix $\tau$ such that  $\tau_{ij}=v_i \cdot v_j$, where the
dot product is taken in the magic basis with no complex conjugation: $v_i
\cdot v_j \equiv \sum_k \bra{e_k}v_i\rangle\bra{e_k}v_j\rangle$.
Consider an arbitrary pure state $\ket{\psi}$ that can be written in the
form $\ket{\psi}=a\ket{v_1}+b\ket{v_2}$.  If $\ket{\psi}$ is expressed as
a 4-vector in the magic basis, we can rewrite
Equation~(\ref{concurrence}) as $C(\psi)=|\psi\cdot\psi|$, and

\begin{equation} C^2(\psi)=(\psi\cdot\psi)(\psi\cdot\psi)^* = \tr[s^*\tau
s\tau^*], \label{tr} \end{equation} where
$s=\mtx{c}{a\\b}\mtx{cc}{a&b}^*$ is the density matrix of $\ket{\psi}$ in
the ($v_1$,$v_2$)-basis.

Let us define the function
\begin{equation}f(\omega)=
\tr[\omega^*\tau\omega\tau^*]\label{f}\end{equation}
for any density matrix $\omega$ expressed in the ($v_1$,$v_2$)-basis.
From Eq.~(\ref{tr}),  $f(\omega)=C^2(\omega)$ if $\omega$ represents a
pure state.  Now, $\omega$ is a $2\times2$ density matrix, and as such
can be written as a real linear combination of Pauli matrices:
$\omega=\frac{1}{2}(I+\vec{r}\cdot\vec{\sigma})$ where
$r_j=\tr[\sigma_j\omega]$.  Substituting this form into Eq.~(\ref{f})
gives us an expression

\begin{equation}
f(\omega)=\frac{1}{4}\tr[\tau^*\tau]+\sum_jr_jL_j+\sum_{i,j}r_ir_jM_{ij}
\end{equation} with

\begin{equation}L_j=\frac{1}{2}\tr[\sigma_j\tau^*\tau]\end{equation} and
\begin{equation}M_{ij}=
\frac{1}{4}\tr[\sigma_i^*\tau\sigma_j\tau^*].\end{equation}
Thus $f$ is defined on the surface and interior of a unit  sphere in
three dimensions, the domain of $\vec{r}$.

$M$ is a real, symmetric matrix with eigenvalues $\pm\frac{1}{2}|\det
\tau|$ and $\frac{1}{4}\tr[\tau^*\tau]$, and $L$ is the eigenvector of
$M$ corresponding to the eigenvalue $\frac{1}{4}\tr[\tau^*\tau]$.  Since
$M$ has two positive eigenvalues and one negative eigenvalue, $f(\omega)$
is convex along two directions and concave along a third.   For the
purpose of this proof, we would like to have a function that is equal to
$f(s)$ for pure states $s$, but convex in all directions.  With this in
mind we define

\begin{equation} g(\omega)=f(\omega)+\frac{1}{2}|\det
\tau|\left(|\vec{r}|^2-1\right), \label{g} \end{equation} which is
identical to $f$ for pure states ($|\vec{r}|=1$). The extra term added to
$f(\omega)$ in effect adds a constant to $f$ and a multiple of the
identity matrix to $M$.  If we define a matrix
\begin{equation}N=M+\frac{1}{2}|\det\tau|I \label{N} \end{equation} and a
constant
\begin{equation}K=
\frac{1}{4}\tr[\tau^*\tau]+\frac{1}{2}|\det\tau|\end{equation}
then we can write

\begin{equation}g(\omega)=
K+\sum_jr_jL_j+\sum_{i,j}r_ir_jN_{ij}.\end{equation}

The added term in Eq.~(\ref{N}) makes all the eigenvalues of $N$
non-negative, one of them being zero.  Thus $g$ is a convex function.
Since $L$ is an eigenvector of $N$ associated with a positive eigenvalue,
and is orthogonal to the eigenvector with zero eigenvalue, the function
$g$ is constant along the latter direction.  We can imagine the function
$g$ (suppressing one dimension)  as a sheet of paper curved  upward into
a parabolic shape; it achieves its minimum value along a straight line.
Moreover, one can show by direct calculation that the minimum value of $g$
is zero.  In the sphere where $\vec{r}$ takes its values, it is
interesting to imagine the surfaces along which $g$ is constant.  These
surfaces appear as  co-axial cylinders with elliptical cross-sections.
(The common axis of the cylinders is the line  along which $g = 0$.)  The
mixed state $\rho$ that we are considering lies on one of these cylinders
and can be decomposed into two pure states lying on the  same cylinder,
that is, having the same value of $g$.   The next two paragraphs show
that no other decomposition of $\rho$ has a smaller average entanglement
than this one.

Any decomposition of $\rho$ into pure states can be viewed as a collection
of weighted points on the surface of the sphere,   such that the ``center
of mass'' of these points is the point representing $\rho$.  The average
entanglement of such an ensemble is the average of ${\cal
E}(\sqrt{g(s)})$ over the ensemble, since ${\cal E}(\sqrt{g(s)})$ is
equal to entanglement for pure states $s$.  If we can show that  ${\cal
E}(\sqrt{g(\omega)})$, regarded as a function of $\omega$,  is convex
over the interior of the sphere, then it will follow that this average
cannot be less than ${\cal E}(\sqrt{g(\rho)})$.  But we have just seen
that $\rho$ can be decomposed into two pure states $s$ for which  $g(s)$
is the same as $g(\rho)$, so this will prove that the entanglement  of
$\rho$ is {\em equal} to ${\cal E}(\sqrt{g(\rho)})$.

In fact it is not hard to prove the desired convexity.  The function
$g(\omega)$ is parabolic with minimum value zero.  Its square root is
therefore a kind of cone and is also convex.  The function ${\cal E}(x)$
is a convex and monotonically increasing function of $x$.  It follows,
then, from the transitive property of convex functions \cite{transitive}
that ${\cal E}(\sqrt{g(\omega)})$ is a convex function of $\omega$.

We have thus found the entanglement of $\rho$ and need only express it in
a simpler form.  Replacing $\omega$ with $\rho$ in Eq.~(\ref{f}) and
using the fact that $\rho$ is diagonal in the $(v_1, v_2)$-basis, we
obtain

\begin{equation} f(\rho) = \tr (R^2) = \lambda_1^2 + \lambda_2^2 ,
\end{equation} where $\lambda_1$ and $\lambda_2$ are the non-zero
eigenvalues of $R$ (Eq.~(\ref{R})). Similarly, one finds that for the
other term in Eq.~(\ref{g}),

\begin{equation} \frac{1}{2}|\det \tau|\left(|\vec{r}|^2-1\right) =
-2\lambda_1\lambda_2, \end{equation} so that $g(\rho) = \lambda_1^2 +
\lambda_2^2 -2\lambda_1\lambda_2$. Taking the square root, we arrive at
the result

\begin{equation} E(\rho)={\cal E}(c); \hskip 2pc c = |\lambda_1 -
\lambda_2|.\label{E2} \end{equation} The expression (\ref{E2}) is
equivalent to Eq.~(\ref{E4}) for  the case of two non-zero eigenvalues.
This completes the proof of the theorem.

Although we have proved our result only for density matrices with just
two non-zero eigenvalues, we can report three pieces of evidence
suggesting that the formula (\ref{E4})  may hold quite generally for a
system of two qubits.   \begin{enumerate} \item For a mixture of Bell
states, mixed with probabilities $p_1, \ldots, p_4$, Bennett {\em et al.}
\cite{formation} have shown that the entanglement  is equal to ${\cal
E}(c)$, with  $c$ given by max $(0, 2p_{max} -1)$.  But in this case $R$
is equal to $\rho$, so that our expression is equal to theirs.  Thus our
formula applies also to this class of density matrices, most of which are
not covered by the above theorem.

\item Peres \cite{Peres} and Horodecki {\em et al.} \cite{Horodecki}
have provided a test, based on partial transposition, for determining
whether a given state of two qubits has zero or  non-zero $E$.  We have
applied both the Peres-Horodecki test and our own formula to   several
thousand randomly chosen density matrices and have found agreement
between  them in every case.  That is, Eq.~(\ref{E4}) gave $E > 0$ if and
only if  the Peres-Horodecki test indicated the presence of entanglement,
which happened in roughly one-third of the cases.

\item For each of twenty-five randomly chosen density matrices with
non-zero entanglement, we have explored the space of all decompositions
of the density matrix into pure states, limiting ourselves to ensembles
of four states.  (The example of Bell mixtures \cite{formation} suggests
that four-state ensembles may be  sufficient.)  In each case, the result
of numerically minimizing the average entanglement of the ensemble agrees
with the result predicted by our formula.

\end{enumerate}

If the formula turns out to be correct for all states, it will
considerably simplify studies of entanglement.  Questions such as whether
the ``distillable entanglement'' is equal to the entanglement of
formation \cite{distill,formation}, that is, whether one can extract as
much entanglement as one puts into the state, will presumably be easier
to answer if there is an explicit formula for the latter quantity. It is
also conceivable that our result can be generalized to systems with
larger state-spaces, such as an entangled pair of $n$-level atoms, though
it is not clear whether there is any structure in such spaces that would
play  quite the same role that the magic basis plays in the two-qubit
case.  In imagining possible generalizations, it is interesting to note
that the form of $R$ has  much in common with the ``mixed-state
fidelity'' \cite{Bures} of Bures, Uhlmann, and Jozsa, which is in no way
special to two-qubit systems.

We would like to thank Charles Bennett, David DiVincenzo and John Smolin
for helpful and stimulating discussions.

\end{document}